# Rectangular Parallelepiped Vibration in Plane Strain State

Jerzy Hańćkowiak
Department of Mechanical Engineering, University of Zielona Góra,
Zielona Góra, Poland
j.hanckowiak@ibmp.uz.zgora.pl

In this paper we present a vibration spectrum of a **homogenous parallelepiped** (HP) under the action of volume and surface forces resulting from the exponent displacements entering the Fourier transforms. Vibration under the action of axial surface tractions and the free vibration are described separately. A relationship between the high frequency vibration and **boundary conditions** (BC) is also considered.

## 1. INTRODUCTION

The catastrophic events in Paris and Moscow are commonly blamed on bad workmanship or improper materials. This reminds us in some sense of other catastrophic events at the beginning of XX century for which oversimplified (engineering) description of structural elements was held responsible, see A.Krilov introduction to Mushelishvili (1949). Hence, as a starting point of the present paper we chose the classical elastodynamical equations

$$\mu\nabla^2\vec{u} + (\lambda+\mu)\nabla(\nabla\cdot\vec{u}) + \rho\cdot(\vec{F}-\ddot{\vec{u}}) = 0$$

(1.1)

where $\mu,\lambda$ - are Lame's elastic constants, $\rho$ - the volume density of HP material, $\rho\cdot\vec{F}$ - the volume force.

One of purposes of the paper is to generalize with the help of Eqs (1.1) existing engineering theories of rectangular beams like Euler-Bernoulli, Timoshenko or even Levinson's new rectangular beam theory, Han et al. (1999), Levinson (1981). We assumed a body in the shape of parallelepiped in the most general plane strain state (PSS) defined in Sec.2.

In the paper we describe the vibration of HP in PSS subjected to surface forces (tractions) acting only upon its bases (ends) and free vibrations.

Our approach is based on *a semi-inverse method* which specifies a family of elementary displacements of HP in PSS by means of which we reproduce the fields of external forces acting upon the HP. *The volume force field* is directly reproduced by means of Eqs (1.1) with unknown $\rho \cdot \vec{F}$. The surface traction field $\vec{t}$ is reproduced by means of Cauchy's formula

$$(t_i(x,y,z,t) = \sigma_{ik}(x,y,z,t)n_k(x,y,z,t))_j \Leftrightarrow (\vec{t}(x,y,z,t) = \sigma(x,y,z,t) \cdot \vec{n}(x,y,z,t))_j$$
(1.2)

with components of stress tensor $\sigma$ and normal vector $\vec{n}$, where *(x,y,z)* are components of an arbitrary point *P* upon the surface of HP (*stress BC*).

At each point of the body under consideration there is a one-to one correspondence between the states of stress and strain expressed by a generalized version of Hooke's law

$$(\sigma_{kl})_j = 2\mu_j(\varepsilon_{kl})_j + \delta_{kl}\lambda_j(\varepsilon_{rr})_j, \quad j = 1,...,p, \quad k,l,r = 1,2,3.$$
(1.3)

where the components of strain tensor $\varepsilon$ are expressed by the components of the displacement vector $\vec{u}$ as follows

$$\varepsilon_{kl} = \frac{1}{2}(u_{k,l} + u_{l,k})$$

In the following we use the notation:

$$\vec{u}(x,y,z,t) = (u_1(x,y,z,t), u_2(x,y,z,t), u_3(x,y,z,t)) \equiv (u,v,w))$$
(1.4)

## 1.1. ACCEPTED CONVENTIONS

We use Einstein's convention in which the repeated index is summed over, if the same index is absent on the other side of an equation. The expression: "wave vector *k*" means only that *k* can also take negative values. To stress that we are concerned exclusively with small vibration, at certain components of the stress vector $\vec{t}$ we omit appropriate components of the normal vector $\vec{n}$.

## 2. PLANE STRAIN STATE (PSS). SOURCE-FREE DISPLACEMENTS

The deformation of a body is described as plane strain if the displacement vector $\vec{u}$ of any point is parallel to a certain plane called *the deformation plane* and is independent of the distance of the point under consideration to this plane, Amenzade (1979). We assume that the deformation plane is identical to plane *OXZ*, Fig.1.

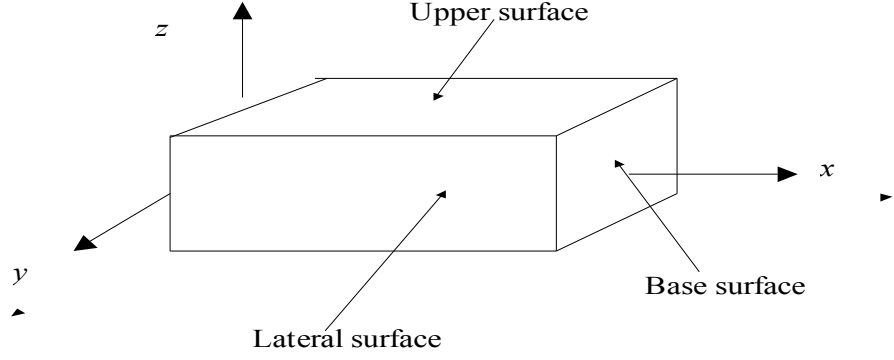

Figure 1. Undeformed homogenous parallelepiped

For a body in PSS, the strain and stress tensors are

$$\varepsilon = \begin{pmatrix} \varepsilon_{xx}, & 0, & \varepsilon_{xz} \\ 0, & 0, & 0 \\ \varepsilon_{zx}, & 0, & \varepsilon_{zz} \end{pmatrix} \equiv \begin{pmatrix} \varepsilon_{11}, & 0, & \varepsilon_{13} \\ 0, & 0, & 0 \\ \varepsilon_{31}, & 0, & \varepsilon_{33} \end{pmatrix}$$

$$\sigma = \begin{pmatrix} \sigma_{xx}, & 0, & \sigma_{xz} \\ 0, & \sigma_{yy}, & 0 \\ \sigma_{zx}, & 0, & \sigma_{zz} \end{pmatrix} \equiv \begin{pmatrix} \sigma_{11}, & 0, & \sigma_{13} \\ 0, & \sigma_{22}, & 0 \\ \sigma_{31}, & 0, & \sigma_{33} \end{pmatrix}$$

(2.1)

Of course, in PSS all components of these tensors, not excluding $\sigma_{yy}$, are $y$-independent, where the variable $y$ describes the distance of a point $P$ to the deformation plane $OXZ$. In PSS, using Hooke's law (1.3),

$$\sigma_{yy} = \lambda(\varepsilon_{xx} + \varepsilon_{zz}) = \frac{\lambda}{2(\lambda + \mu)}(\sigma_{xx} + \sigma_{zz})$$

(2.2)

This relation among components of the stress tensor in PSS means that in the case of free lateral surfaces of HP, $\sigma_{yy} = 0$, and this leads to the source-free displacement field – a property with many implications. In this case, which we assume throughout the paper, the PSS is simultaneously the plane stress state, because by definition, *the plane stress state* is characterized by the following stress tensor:

$$\sigma = \begin{pmatrix} \sigma_{xx}, & 0, & \sigma_{xz} \\ 0, & 0, & 0 \\ \sigma_{zx}, & 0, & \sigma_{zz} \end{pmatrix}$$

(2.3)

So, from (1.3), (2.2), (2.3) and PSS we get

$$\theta \equiv (\varepsilon_{xx} + \varepsilon_{yy} + \varepsilon_{zz}) = (\varepsilon_{xx} + \varepsilon_{zz}) = u_{,x} + w_{,z} = 0 \tag{2.4}$$

In other words, we consider *the source-free* displacement field $\vec{u}$. A further implication of the assumptions accepted is the possibility of separating of Eqs (1.1): we get the following equations for components

$$\mu \cdot (u_{,xx} + u_{,zz}) + \rho \cdot (X - \ddot{u}) = 0$$

$$\mu (w_{,xx} + w_{,zz}) + \rho \cdot (Z - \ddot{w}) = 0 \tag{2.5}$$

The 3D character of our problem will be manifested only through the boundary conditions (BC), which simultaneously connect together the non vanishing components *u,w* of the displacement vector.

For comparison, we write out the stress tensor for the Saint Venant problem describing torsion and bending of a prismatic beam acted on by tractions at the end faces

$$\sigma = \begin{pmatrix} \sigma_{xx}, \sigma_{xy}, \sigma_{xz} \\ \sigma_{yx}, 0, 0 \\ \sigma_{zx}, 0, 0 \end{pmatrix} \tag{2.6}$$

Huber (1954; vol. 1, page 284), and the stress tensor for classical, engineering theory of thin plates, loaded by the tractions acting upon the upper, bottom and lateral faces

$$\sigma \cong \begin{pmatrix} \sigma_{xx}, \sigma_{xy}, 0 \\ \sigma_{yx}, \sigma_{yy}, 0 \\ 0, 0, 0 \end{pmatrix} \tag{2.7}$$

where only quantities of the first order were taken into account. The left components of the stress tensors appear in the form of inhomogeneous terms or coefficients - in derived, simplified (engineering) equations, Kączkowski (2000; page-24) and can be treated as *sources* of the field under consideration.

### 3. PARTICULAR DISPLACEMENTS AND LOADINGS

We would like to show that by means of exponential functions entering the Fourier transform and by means of superposition principle one can construct elementary displacements which satisfy Eqs (2.4-5) or equivalently Eqs (1.1). Moreover, parameters entering displacements can be chosen in such a way that the surface tractions appear only upon the bases of HP. For the particular choice of parameters under consideration, the volume forces also disappear.

Following Romanów (1995) we do not specify the thickness of a body.

We consider the source-free displacement field

$$u^{\pm}(x,z,t) = \frac{i}{k} D^{\pm} \Pi^{\pm}_{,z}(z) \exp(ik \cdot x) \cdot \exp(\pm i\omega \cdot t)$$

$$w^{\pm}(x,z,t) = \{C^{\pm} + D^{\pm} \cdot \Pi^{\pm}(z)\} \exp(ik \cdot x) \cdot \exp(\pm i\omega \cdot t)$$

(3.1)

where $k$ – the wave vector, $\omega$ - the angular frequency. (3.1) satisfies (2.4) without any restriction of parameters and any specification of the function $\Pi$. The upper indices $\pm$ mean that two independent solutions related to the sign in front of the angular frequency $\omega$ are considered.

Constants $C^{\pm}, D^{\pm}, H^{\pm},...$, in the above formulas, may in fact depend on additional variables such as $k, \omega, n$ which increases the number of possible solutions. In that case we call them *dispersion functions*.

3.1. VOLUME FORCE SPECIFICATION

Substituting (3.1) in Eqs (2.5), the components of the volume force field are obtained. For a particular choice of the function

$$\Pi^{\pm}(z) = H^{\pm} \exp(\sqrt{c} \cdot z) + K^{\pm} \exp(-\sqrt{c} \cdot z)$$

(3.2)

with constant

$$c \equiv k^2 - \left(\frac{\omega}{c_T}\right)^2; \quad c_T = \sqrt{\mu/\rho}$$

(3.3)

where $c_T$ – the phase velocity of transverse waves in an unbounded medium, the volume force $\rho \cdot \vec{F}$ has a vanishing horizontal component and does not depend on the variable *z*. These rather desired properties of the volume forces are related to the (3.3) "synchronization" of the *z, x* and *t*-dependences of displacements (3.1-2). No *amplitudes* (constants $C, D, H, K$) enter (3.3).

By direct examination of Eqs (2.5) it can be shown that displacements (3.1) with *C=0* and (3.2) correspond to the volume-free force, Sec.4.5.

3.2. SHEAR-FREE BC ON TOP AND BOTTOM SURFACES OF HP ($z = \pm h$).

This condition - upon the shear components of the stress vector $\vec{t}$ upon the top and bottom surfaces of HP in PSS - leads to equation

$$t_1 = \sigma_{13} n_3 = 2\mu\varepsilon_{13} n_3 \Rightarrow (u_{,z} + w_{,x})|_{z=\pm h} = 0$$

(3.4)

This means that the above shear free BC for HP lead to vanishing of the shear strains upon $z = \pm h$. From (3.4) and (3.1-3) **which involve (2.4)** we get equation

$$D\Pi(\pm h)\left[2 - \left(\frac{\omega}{c_T \cdot k}\right)^2\right] + C = 0$$

(3.5)

These two equations involving the four amplitudes $C, D, H, K$ and the two parameters $\omega, k$ can be satisfied in different ways considered below.

3.3. NORMAL-FREE BC ON TOP AND BOTTOM SURFACES OF HP ($z = \pm h$)

This condition upon the normal components of the stress vector $\vec{t}$ (together with (2.4)) leads to equations

$$t_3 = \sigma_{33} n_3 = (2\mu\varepsilon_{33} + \lambda\varepsilon_{kk})n_3 = 2\mu\varepsilon_{33} n_3 \Rightarrow w_{,z}\big|_{z=\pm h} = 0$$

(3.6)

which also mean that the normal strains at $z = \pm h$ vanish. From (3.6), two equations result

$$H = K\exp(\pm\sqrt{c}\cdot 2h)$$

(3.7)

which if treated as equations upon the constants $H, K$ lead to the following restrictions for the values of the constant (3.3)

$$c \equiv c_n = -\left(n\frac{\pi}{2h}\right)^2; \quad n = 0, \pm 1, \pm 2, \ldots$$

(3.8)

In this way, for odd $n$,

$$H = -K$$

(3.9)

and, for $0$ and even $n$,

$$H = K$$

(3.10)

Hence and from (3.2) the corresponding expressions for functions $\Pi(z)$ result:
In the case of (3.9)

$$\Pi(z) \equiv \Pi_{2n+1}(z) = -2iK_{2n+1}\sin\left(\frac{(2n+1)\pi}{2h}z\right)$$

(3.11)

and in the case of (3.10)

$$\Pi(z) \equiv \Pi_{2n}(z) = 2K_{2n}\cos\left(\frac{n\pi}{h}z\right)$$

where in both cases $n = 0, \pm 1, \pm 2, \ldots$ For both cases, we have dispersion relations

$$\omega^2 \equiv \omega_n^2(k) = c_T^2\left(k^2 + \left[\frac{n\pi}{2h}\right]^2\right)$$

(3.12)

in which the discrete index $n = 0,\pm1,\pm2,...$ From (3.11) results

$$\Pi_{2n+1}(\pm h) = -2iK_{2n+1}\sin\left(\pm\frac{(2n+1)\pi}{2}\right) = \mp 2iK_{2n+1}\sin\left(\frac{(2n+1)\pi}{2}\right) = \mp(-1)^n 2iK_{2n+1}$$

(3.13)

$$\Pi_{2n}(\pm h) = 2K_{2n}\cos(\pm n\pi) = 2K_{2n}(-1)^n$$

This means that in the case when the normal tractions disappear upon the top and bottom surfaces of HP in PSS, the functions $\Pi(z)$ do not vanish upon these surfaces. As a result of this property and equation, $C=0$, the dispersion relations (3.16) can be derived.

3.4. VOLUME AND HP GENERATORS –FREE BC. AN ALTERNATIVE

As we have seen above, the absence of shear tractions upon the bottom and top surfaces of HP and the free-lateral BC in PSS leads to Eqs (3.5), which in the case of $C=0$ (absence of volume forces; Sec.4) give the following alternative:

$$D\Pi(\pm h) = 0 \quad or \quad 2 - \left(\frac{\omega}{c_T \cdot k}\right)^2 = 0$$

(3.14)

The above alternative together with shear-free BC on the bottom and top surfaces and lateral-free BC guarantee that the volume forces, related to the displacements (3.1), disappear.

In the case of the normal-free BC on the bottom and top surfaces and lateral -free BC (Sec.3.3) the first term of the alternative (3.14) cannot occur because of (3.13). From the second term of the above alternative the following dispersion relation implies

$$\omega^2 = 2c_T^2 k^2$$

(3.15)

They together with (3.3) and (3.8) lead to the following spectrum of possible vibration

$$k^2 \equiv k_n^2 = \left(n\frac{\pi}{2h}\right)^2 \quad and \quad \omega_n^2 \equiv \omega_n^2(k_n) = 2c_T^2\left(n\frac{\pi}{2h}\right)^2$$

(3.16)

where $n = 0,\pm1,\pm2...,$ We have to remember that for odd $n$ the function $\Pi_n(z)$ is given by (3.11) and for even $n$ is given by (3.12).

3.5. WEAKENED BC ON TOP AND BOTTOM SURFACES OF HP

Displacements (3.1), for an arbitrary choice of functions $\Pi(z)$, satisfy the traction-free BC on the lateral surfaces of the HP (Eqs. (2.4)) in all cases considered. It turns out that additional conditions assuming the traction-free BC on the top, bottom and lateral surfaces lead to very high frequency vibration in which the wavelength of vibration modes may be comparable to the thickness dimension of the HP, see (3.16). In the face of such results we would like to examine some weakening of BC. **We base upon intuition that vibration of HP made of thin and hard materials do not depend on the normal components of the tractions acting on the top and bottom surfaces, when their absolute values are approximately equal at both sides of the HP**.

There is an easy way of checking that the above intuition holds the static case. A sheet of paper deflecting under the action of the horizontal tractions acting on their ends will not change its deflection under the influence of an arbitrary compensated field of the normal tractions acting upon the top and bottom surfaces.

Taking an arbitrary field of displacements we cannot be sure that they correspond to the above commonly compensated normal tractions. However, in the case of a slowly changing field and thin HP this occurs in every case.

Let us examine such a possibility, for the alternative (3.14). Using the first term of the alternative (3.14), we get from (3.2)

$$\Pi(\pm h) = H\exp(\pm\sqrt{c}\cdot h) + K\exp(\mp\sqrt{c}\cdot h) = 0$$

(3.17)

where the upper indices have been intentionally omitted to avoid confusion with signs in front of the square root of *c*. Considering Eqs (3.17) as a system of two homogenous equations upon the two unknowns *H,K*, we get the following set of possible values for the constant *c*:

$$\sqrt{c} \equiv \sqrt{c_n} = i\frac{n\pi}{2h}; \quad n = 0, \pm 1, \pm 2, ...$$

(3.18)

These are exactly the same restrictions as (3.8), but now derived by means of different BC. They lead however, together with (3.17), to different expressions upon functions $\Pi(z)$

$$\Pi_{2n+1}(z) = 2K_{2n+1}\cos\left(\frac{(2n+1)\pi}{2h}z\right)$$

(3.19)

$$\Pi_{2n}(z) = -2iK_{2n}\sin\left(\frac{2n\pi}{2h}z\right)$$

These formulae are similar to (3.11), but now instead of dispersion relations (3.15-16) we get, via (3.18) and (3.3),

$$\omega^2 \equiv \omega_n^2(k) = c_T^2\left(k^2 + \left[n\frac{\pi}{2h}\right]^2\right)$$

(3.20)

in which the wave vector $k$ is not restricted at all. It is interesting to notice that we get exactly the same formula for the angular frequency using only the normal-free BC on $z = \pm h$. Only when we simultaneously use the normal-free and shear-free BC (no tractions on $z = \pm h$), do we get formulas (3.16).

Using the second term of the alternative (3.14) we get the dispersion relation (3.15), and from (3.3)

$$c = -k^2$$

(3.21)

Hence and from (3.2)

$$\Pi^{\pm}(z) = H^{\pm}\exp(ik \cdot z) + K^{\pm}\exp(-ik \cdot z)$$

(3.22)

The functions (3.22), for an appropriate choice of $k$, see (4.12), may be slowly varying functions along the thickness of HP. This makes it possible to satisfy the proposed weakening of BC. In the first case, in which functions $\Pi_n(z)$ are given by (3.19) and $h$ is a small quantity, a small deviation of HP from the ideal shape of the parallelepiped may cause lack of compensation of the normal tractions on the opposite surfaces.

Restrictions upon the vector $k$ can be obtained if instead of the rejected condition (3.6) we use other more familiar conditions like vanishing of tractions or their components upon the bases (end faces) of HP. But for that purpose we have to use more general solutions than (3.1) by using the superposition principle.

3.6. PLAIN CROSS-SECTIONS

Choosing in (3.22)

$$H^{\pm} = K^{\pm}$$

(3.23)

we get

$$\Pi^{\pm}(z) = 2K^{\pm}\cos(k \cdot z)$$

(3.24)

In this case, for thin HP and appropriate $k$, the displacements (3.1) are given by

$$u^{\pm}(x,z,t) \cong -2iD^{\pm}K^{\pm} \cdot z \cdot \exp(ik \cdot x)\exp(\mp i\omega(k) \cdot t)$$

$$w^{\pm}(x,z,t) \cong \{C^{\pm} + 2D^{\pm}K^{\pm}\}\exp(ik \cdot x)\exp(\mp i\omega(k) \cdot t)$$

(3.25)

where the angular frequency $\omega$ is given by (3.15).

# 4. MORE GENERAL DISPLACEMENTS AND LOADINGS

More general solutions to Eqs (1.1) one can be constructed with the help of the superposition principle using the elementary solutions (3.1) or (3.25). Let us express the displacement vector as

$$\vec{u}(x,z,t) = \int \vec{U}(k,\omega,z) e^{ik\cdot x} e^{i\omega\cdot t} dk d\omega$$

(4.1)

where

$$\vec{U}_n^{\pm} = \left( U_n^{\pm}, 0, W_n^{\pm} \right)$$

(4.2)

$$U^{\pm}(k,\omega,z) = \frac{i}{k} D^{\pm}(k,\omega) \Pi_{,z}^{\pm}(k,\omega,z), \quad W^{\pm}(k,\omega,z) = \left\{ C^{\pm}(k,\omega) + D^{\pm}(k,\omega) \cdot \Pi^{\pm}(k,\omega,z) \right\}$$

The previous constants $C, D, H, K$ of solutions (3.1) or (3.25) are treated here as the unknown *spectral functions* depending on the wave vector $k$ and the angular frequency $\omega$. The functions $\Pi_n^{\pm}(k,\omega,z)$ are given by formulae (3.2) and (3.3). In this case the volume force $\rho \cdot \vec{F}$ has a vanishing horizontal component and does not depend on the variable *z*, see below. The tractions on the lateral surfaces of HP also vanish.

Postulating for the *vector spectral function* $\vec{U}(k,\omega,z)$ the following form (comb filter in the variable $\omega$):

$$\vec{U}(k,\omega,z) = \sum_n \vec{U}_n^+(k,z) \delta(\omega + \omega_n(k)) + \vec{U}_n^-(k,z) \delta(\omega - \omega_n(k))$$

(4.3)

we admit only a dependence of the angular frequency $\omega$ on the wave vector $k$ and the additional, discrete index *n*. This takes place for the normal-free BC on $z = \pm h$, Sec.3.3. Additional BC imposed on the top and bottom surface of HP (sic!) or at the bases of it restrict also values of the wave vector *k*, see (3.16) and below.

4.1. TRACTIONS AT BASES OF HP

Tractions acting on the bases (ends) of HP can be calculated by means of the following patterns: for the normal component (*OX*-projection):

$$t_1(x,z,t)|_{x=0,L} = \sigma_{11}(x,z,t) n_1(x,y,z,t)|_{x=0,L} = 2\mu \cdot u_{,x} n_1(x,y,z,t)|_{x=0,L} =$$
$$2i\mu \int U(k,\omega,z) \cdot k e^{ik\cdot x} e^{i\omega\cdot t} dk d\omega \cdot n_1(x,y,z,t)|_{x=0,L} \cong$$
$$\pm 2\mu \sum_{n,\pm} \int \left( \Pi_{n,z}^{\pm}(k,z) D_n^{\pm}(k) \exp[\mp i\omega_n(k) t] \right) e^{ik\cdot x} dk|_{x=0,L}$$

(4.4)

where + before $2\mu$ corresponds to the traction at *x=0* and "–" at *x=L*.

For the shear component (*OZ*-projection)

$$t_3(x,z,t)|_{x=0,L} = \sigma_{31} n_1 |_{x=0,L} = 2\mu\varepsilon_{31} n_1 = \mu(u_{,z} + w_{,x}) n_1 |_{x=0,L} =$$
$$\cong i\mu \sum_{n,\pm} \int D_n^\pm(k) \Pi_n^\pm(k,z) \left\{ \frac{c_n}{k} + k \right\} \exp[\mp i\omega_n(k) \cdot t] \cdot e^{ikx} |_{x=0,L} \, dk$$

(4.5)

The symbol $\cong$ has been used to express fact that for small vibration we can write

$$n_1(x,y,z,t) \cong 1$$

(4.6)

The constants $c_n$ may in fact not depend on $n$ as in the case of (3.21) when (4.5) is identically equal to zero. The vertical component $t_3$ of the stress vector $\vec{t}$ is also equal to zero in the case of volume and generator-free BC, Sec. 3.4. This component is not equal to zero in the case of the dispersion relation (3.20) which corresponds to the first term of alternative (3.14).

If $C \neq 0$ there are also volume forces acting on the body.

## 4.2. TRACTION-FREE BC ON TOP, BOTTOM AND LATERAL SURFACES. AXIAL TRACTIONS

In this case we have spectral relations (3.15-16). This means that spectral functions have the *filtering property of the Dirac delta-function*:

$$D_n^\pm(k)\Pi_n^\pm(k,z) = D_n^\pm \Pi_n^\pm(z)\delta(k - n\pi/2h); \quad n = 0, \pm 1, \pm 2, \ldots$$

(4.7)

As a consequence of this filtering property the shear tractions on the ends of HP are vanishing, see (3.8) and the expression in the curled brackets in (4.5). This is a **surprising result**: the traction-free BC on the top, bottom and lateral surfaces of HP enforced also the shear-free BC on the bases of HP. In other words, formulas (4.1-3) with spectral functions (4.7) and $C_n^\pm(k) = 0$, correspond to forced vibration of HP in PSS under *axial tractions* (4.4) only.

## 4.3. ALMOST FREE VIBRATION

Let us notice that for the antisymmetric spectral functions

$$D_n^\pm(k)\Pi_n^\pm(k,z) = -D_n^\pm(-k)\Pi_n^\pm(-k,z)$$

(4.8)

instead of (4.4-5), we get the formulas:

$$t_1(x,z,t)|_{x=0,L} \cong \pm 4i\mu \sum_{n,\pm} \int_0^\infty \Pi_{n,z}^\pm(k,z) D_n^\pm(k) \exp[\mp i\omega_n(k)t] \sin(kx) dk |_{x=0,L}$$

(4.9)

and

$$t_3(x,z,t)\vert_{x=0,L} =$$

$$-2i\mu \sum_{n,\pm} \int_0^\infty D_n^\pm(k)\Pi_n^\pm(k,z)\left\{\frac{c_n}{k}+k\right\}\exp[\mp i\omega_n(k)\cdot t]\sin(kx)n_1(x,z,t)\vert_{x=0,L}\, dk$$

(4.10)

where in the last formula the *x*-component of the normal vector $\vec{n}$ has been left to show that the conclusion below does not depend on the approximation (4.6). Taking into account the arbitrariness of *k* we postulate the comb filter property for the product of the spectral functions:

$$D_n^\pm(k)\Pi_n^\pm(k,z) = \sum_l D_{nl}^\pm \Pi_{nl}^\pm(z)\delta(k-k_l); \quad k \geq 0$$

(4.11)

In the case of (3.20) and (4.11), the terms with angular frequencies given by

$$\omega_n^2(k_l) = c_T^2\left(k_l^2 + \left[\frac{n\pi}{h}\right]^2\right)$$

(4.12)

give contributions to (4.10). In the case of (3.15)

$$\omega_n^2(k_l) = 2c_T^2 k_l^2$$

(4.13)

and

$$D_{nl}^\pm \Pi_{nl}^\pm(z) = \delta_{nl} D_n^\pm \Pi_n^\pm(z)$$

(4.14)

In both cases

$$k \equiv k_l = \frac{l\cdot\pi}{L}; \quad l = 0,\pm 1,\pm 2\ldots$$

(4.15)

where the length of HP is in the denominator – *L*.

Hence, we come to the following
*Statement*:

Displacements (4.1-3), with discrete spectral functions (4.8) and (4.11), satisfy *almost free* BC.

"Almost free" means that in the weakened BC the normal tractions on the top and bottom surfaces of HP may not vanish but they should be compensated to each other. The reader will remember that due to such BC long wavelength vibration of HP are possible.

To use the weakened BC proposed in Sec. 3.5, we need displacements slowly varying in the *z*-variable, when $z \in <-h,h>$. This takes place for thin HP, at least for the few first *l*, when in the displacements (3.1) the functions $\Pi^\pm(z)$ depend on *k* given by (4.15) but not by (3.16).

4.4. TRULY FREE VIBRATION

If in formulas (4.8-10) for components of the stress vector $\vec{t}$ the spectrum is given by (4.7), see (3.16), then we have case of no tractions on generator surfaces of HP. If additionally

$$\sin\left(\frac{n\pi}{2h}x\right) = \sin\left(\frac{l\pi}{L}x\right)$$

(4.16)

then tractions on the bases of HP also vanish and we have a case of truly free vibration of HP. Unfortunately, in the present approach this case takes place only for very short waves which by definition are much shorter then the overall dimensions of the body, Langley et al. (1998).

4.5. VOLUME FORCES. SINGULARITIES OF THIN HP

The elementary displacements (3.1) are responsible also for presence of the volume forces acting on HP. Choosing (3.3) of the multiplicative constant $c$ regulating the $z$-behavior of theory, the unvanishing component of the volume forces is the vertical one

$$\rho \cdot Z(x,t) \equiv \rho \cdot \ddot{w} - \mu \cdot (w_{,xx} + w_{,zz}) =$$
$$\sum_{n,\pm} \int C_n^{\pm}(k) \{-\rho \cdot \omega_n^2(k) + \mu \cdot k^2\} \exp[\mp i\omega_n(k) \cdot t] \cdot e^{ikx} dk =$$
$$-\mu \sum_{n,\pm} \int C_n^{\pm}(k) \cdot c_n(k) \exp[\mp i\omega_n(k) \cdot t] \cdot e^{ikx} dk$$

(4.17)

where in the last row we have taken into account that the constant $c$ defined in (3.3) may depend on $k$ and $n$, see (3.8), (3.18), (3.21). We examine this component in the case of vanishing tractions on $z = \pm h$ and of course without lateral tractions. We have (3.5), (3.8) and (3.12). We have to consider only even solutions for functions $\Pi_n^{\pm}(z)$ because otherwise, $C_n^{\pm}(k)$ calculated by means of Eqs (3.5) would not be unique. From (3.5), (3.11) and (3.13)

$$C_{2n}^{\pm}(k) = 2(-1)^{n+1} D_{2n}^{\pm}(k) K_{2n}^{\pm}(k) \left[1 - \left(\frac{n \cdot \pi}{h \cdot k}\right)^2\right]; \quad 1 \neq \left(\frac{n \cdot \pi}{h \cdot k}\right)^2$$

(4.18)

The separate terms of the sum (4.17) may take large values, for thin HP ($h \ll 1$). This property is additionally enhanced for long HP ($L \gg 1$) when tractions on the ends of HP vanish, see (4.15). In other words, **for thin HP**, it is much easier to produce harmonic vibrations with the alternative (3.14) and the dispersion relation (3.15) than to produce vibration caused by non vanishing volume force (4.17) with the dispersion relations (3.12).

## 5. FINAL REMARKS

The difference of present approach from Levinson's work (1985) consists in the assumption that HP is in PSS and that we exclusively use the stress BC.

The displacement BC can be deduced from the stress BC. In particular, the engineering BC which depend on used models, see Han et al. (1999), can be deduced.

We carefully examine the influence of the BC used upon the space and time dependence of elementary displacements (3.1). Lack of tractions on the lateral surfaces of HP led to source-free displacements. Lack of tractions on the top and bottom surfaces led to discrete values of the wave vector $k$ and to high frequency vibration (sic!) particularly for thin HP. This phenomenon is probably related to the fact that we consider simultaneously in-plane and out-plane vibration, Hyde et al. (2001). The fact that with the help of weakened BC, see below, one can describe low frequency and low wave vibration **contrasts** with results of engineering theories which show that for

$$\omega > \omega_c = \frac{1}{k}\sqrt{\frac{k'G}{\rho}}$$

(5.1)

the spatial solution only has sinusoidal terms, see Han et al. (1999; 950 page). This may means that sinusoidal solutions obtained in this way can be unstable and particularly in the vacuum and weightlessness conditions only short sinusoidal waves can be observed.

It turns out that restrictions coming from the disappearance of only one component of the stress vector $\bar{t}$ on the top and bottom surfaces of HP did not confine values of the wave vector $k$. This fact and the hypothesis that normal components do not influence deflections of **thin HP** if the top and bottom parts have approximately equal absolute values but opposite senses, allows us to take into account, at the ends (bases) of HP, additional BC which admit lower frequency vibration. In general, these are forced vibration, but after choosing antisymmetric spectral functions (4.8) – the almost free vibration can be obtained.

The vibration under axial tractions acting on the bases of HP are described by formulas (4.1-3) in which the spectral functions have the filtering property of the Dirac delta function (4.7) and the dispersion relations (3.16) take place. In this case boundedness of vibration is reduced to examination of boundedness of the Fourier transforms of the spectral functions, Glisson (1985).

It is also worth noticing that the presence of the $z$-independent component in the vertical vibration ($C \neq 0$), which in engineering description is exclusively assumed, is responsible for the presence of volume forces depending on the thickness and length of HP in a way which is not acceptable. These forces diminish with the growing thickness of HP.

At the end we give some illustrations of elementary displacements (3.1) confining ourselves to their real or imaginary parts, of course. We use the public program "SmallDyn2", see Nusse et al. (1997), with the help of which we draw the direction and displacement fields. In the case of PSS we obtain full information about these fields drawing them in a cross-section parallel to the deformation plane *OXZ*. Such a cross-section is represented by a net of uniformly situated nodes which represent the distinguished particles or cells before deformation. In the case of the direction field, an arrow is going out from every node which sense is changing after one cycle. For the displacement field, the length of arrows is proportional to the length of the displacement vector at a given point.

Let us assume that we have two conjugate solutions (3.1) with *k* and *–k* such that (4.8) takes place. Then from the superposition principle we get a new solution

$$u_n^\pm(x,z,t) = \frac{2i}{k} D^\pm(k) \Pi_{n,z}^\pm(k,z) \cos(kx) \exp(\pm i\omega_n(k)\cdot t)$$

$$w_n^\pm(x,z,t) = 2i \cdot \left\{ C_n^\pm(k) + D_n^\pm(k)\Pi_n^\pm(k,z) \right\} \sin(kx) \exp(\pm i\omega_n(k)\cdot t)$$

(5.1)

In the last equation we took into account the restriction (3.5) from which results $C^\pm(k) = -C^\pm(-k)$. Hence, we get, see (3.11) or (3.24),

$$u = -\frac{C1}{C2} * C3 * \sin(C1*z) * \cos(C2*x)$$

$$w = C3 * \cos(C1*z) * \sin(C2*x) + C4 * \sin(C2*x)$$

(5.2)

Here the "constant" *C3* represents the time-dependent exponent of (5.1). Other constants *C1,2* are related in an obvious way to *c* and *k*. In fact, due to complex number description used in the paper, we can consider trigonometric functions with other initial phases, like in Zhao et al. (2001) where such products are considered as bases functions, for a description of the high frequency vibration.

We illustrate displacements (5.2), for $C1 = C2 = n\pi/2h$, *n=1, h=0.1m*. In all pictures - *C3,* which harmonically depends on the time *t*, is constant and equal to one.

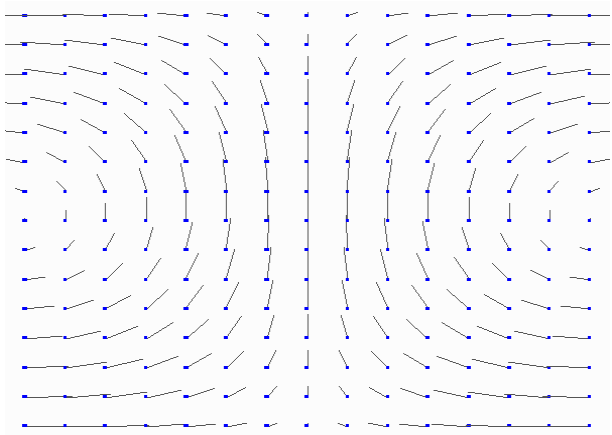

Fig.2. L=2h, h=0.1m
Fig.3. L=3h

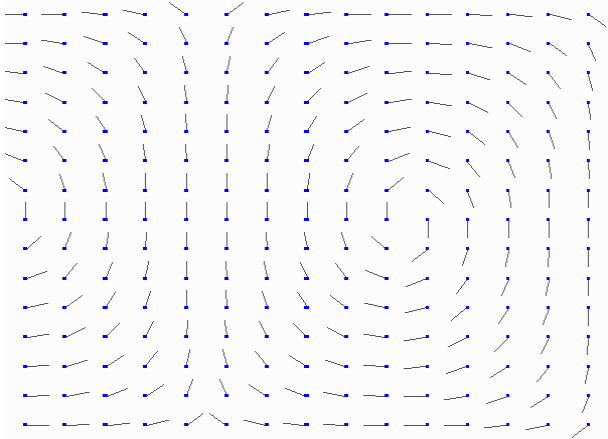 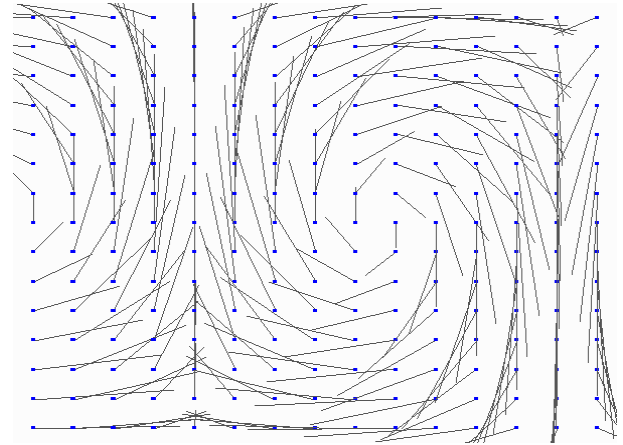

Fig.4. L=3h {direction field}  Fig.5. L=3.3h

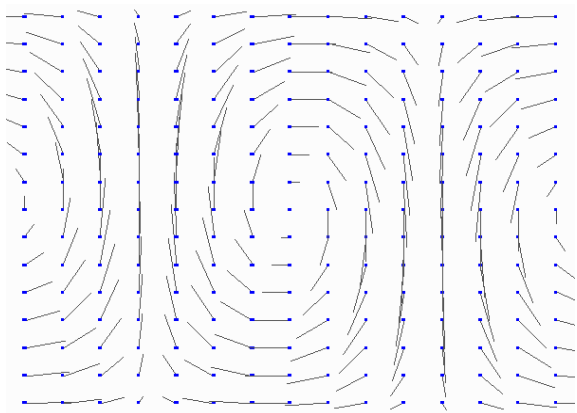 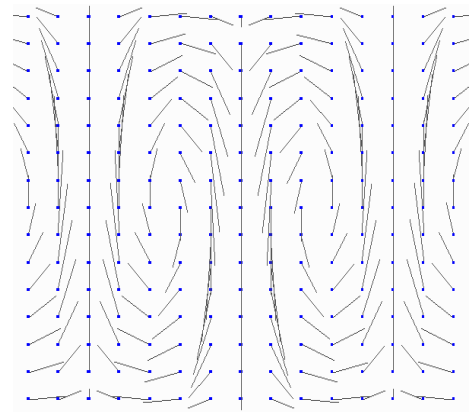

Fig.6. L=4h  Fig.7. L=6h

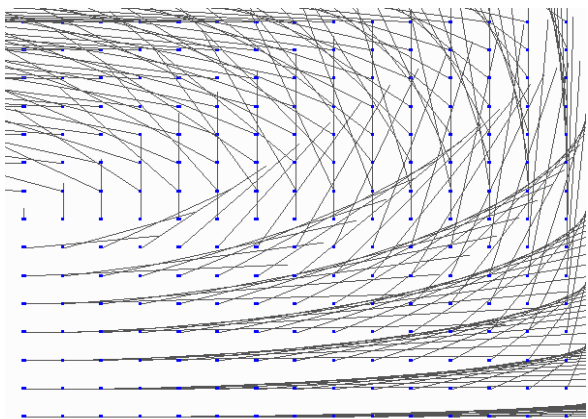

Fig.8. L=h
Fig.9. L=h {direction field}

One can see that the displacements in Figures 3, 5 and 8, which correspond to axial tractions operating on the ends of HP, are relatively big. It is remarkable that in all cases the corresponding direction fields are very regular and clear express the source-free character of the used displacements.